\documentclass[prb,superscriptaddress,amsfonts,twocolumn,showpacs,floatfix]{revtex4-1}
\usepackage{amsmath}
\usepackage{amsfonts}
\usepackage{amssymb}
\usepackage{bm}
\usepackage{txfonts}
\usepackage{tabularx}
\usepackage{multirow}
\usepackage[dvips]{graphicx,color}

\begin{document}
\title{%
Cubic lead perovskite PbMoO$_3$ with anomalous metallic behavior
}

\author{Hiroshi Takatsu}
\affiliation{Department of Energy and Hydrocarbon Chemistry, Graduate School of Engineering, Kyoto University, Kyoto 615-8510, Japan}

\author{Olivier Hernandez}
\affiliation{Institut des Sciences Chimiques de Rennes, UMR CNRS 6226, Universit$\acute{\rm e}$ de Rennes 1, Batiment 10B, Campus de Beaulieu, Rennes F-35042, France}

\author{Wataru Yoshimune}
\affiliation{Department of Energy and Hydrocarbon Chemistry, Graduate School of Engineering, Kyoto University, Kyoto 615-8510, Japan}

\author{Carmelo Prestipino}
\affiliation{Institut des Sciences Chimiques de Rennes, UMR CNRS 6226, Universit$\acute{\rm e}$ de Rennes 1, Batiment 10B, Campus de Beaulieu, Rennes F-35042, France}

\author{Takafumi Yamamoto}
\affiliation{Department of Energy and Hydrocarbon Chemistry, Graduate School of Engineering, Kyoto University, Kyoto 615-8510, Japan}

\author{Cedric Tassel}
\affiliation{Department of Energy and Hydrocarbon Chemistry, Graduate School of Engineering, Kyoto University, Kyoto 615-8510, Japan}

\author{Yoji Kobayashi}
\affiliation{Department of Energy and Hydrocarbon Chemistry, Graduate School of Engineering, Kyoto University, Kyoto 615-8510, Japan}

\author{Dmitry Batuk}
\affiliation{EMAT, University of Antwerp, 2020 Antwerpen, Belgium}

\author{Yuki Shibata}
\affiliation{Department of Energy and Hydrocarbon Chemistry, Graduate School of Engineering, Kyoto University, Kyoto 615-8510, Japan}

\author{Artem M. Abakumov}
\affiliation{EMAT, University of Antwerp, 2020 Antwerpen, Belgium}
\affiliation{Skolkovo Institute of Science and Technology, Nobel str. 3, 143026 Moscow, Russia}

\author{Craig M. Brown}
\affiliation{National Institute of Standards and Technology, Center for Neutron Research Gaithersburg, MD 20899-6102, U.S.A.}

\author{Hiroshi Kageyama}
\affiliation{Department of Energy and Hydrocarbon Chemistry, Graduate School of Engineering, Kyoto University, Kyoto 615-8510, Japan}

\date{\today}

\begin{abstract}
A previously unreported Pb-based perovskite PbMoO$_3$ is obtained by high-pressure and high-temperature synthesis. 
This material crystallizes in the $Pm\bar{3}m$ cubic structure at room temperature, 
making it distinct from typical Pb-based perovskite oxides with a structural distortion.
%
PbMoO$_3$ exhibits a metallic behavior down to 0.1~K
with an unusual $T$-sub linear dependence of the electrical resistivity.
Moreover, a large specific heat is observed at low temperatures accompanied by a peak in $C_P/T^3$ around 10~K,
in marked contrast to the isostructural metallic system SrMoO$_3$. 
These transport and thermal properties for PbMoO$_3$, taking into account anomalously large Pb atomic displacements
detected through diffraction experiments, 
are attributed to a low-energy vibrational mode, associated with incoherent off-centering of lone pair Pb$^{2+}$ cations.
We discuss the unusual behavior of the electrical resistivity in terms of a polaron-like conduction,
mediated by the strong coupling between conduction electrons and optical phonons of the local low-energy vibrational mode.
\end{abstract}


\maketitle

\section{Introduction}
Insulating oxides with a stereochemically active lone pair of Pb$^{2+}$ and Bi$^{3+}$ cations have been 
the subject of long-standing interest due to large remnant polarization and high temperature structural/electronic transitions, as found in Pb(Zr,Ti)O$_3$ 
with a coherent displacement of Zr/Ti atoms~\cite{N.Setter, B.Jaffe}.
A surge of interest in multiferroics over the last decade has led to numerous studies of BiFeO$_3$~\cite{WangScience2003}
targeting applications in ferroelectric non-volatile memory devices with high working temperature. 
More recently, there has been a growing attention in ferroelectric-like structural distortion or transition in 
metallic compounds~\cite{Anderson1965,Sergienko2004,H.SakaiSA2016,KoziiPRL2015,ShiNM2013}.
This is because unprecedented behaviors can appear distinct from conventional metals, 
such as an enhanced thermoelectric property in (Mo,Nb)Te$_2$~\cite{H.SakaiSA2016} and possible odd-parity superconductivity proposed in
Cd$_2$Re$_2$O$_7$ and doped SrTiO$_3$ heterostructures~\cite{KoziiPRL2015}. 

The concept of introducing polar distortion 
in a metallic phase has been recently demonstrated 
in a material of the LiNbO$_3$-structure type, LiOsO$_3$~\cite{ShiNM2013}.
It is discussed that large thermal vibration or incoherent disorder of Li and O ions allows cooperative order upon cooling.
A ferroelectric-type structural phase transition occurs at $T_{\rm s}=140$~K, 
where the metallic conduction associated with Os half-filled $t_{2\rm g}$ states is retained across $T_{\rm s}$.
In this context,
compounds with stereochemically active lone-pair electrons 
can be considered as an alternative path for
designing ferroelectric-like metal or a new class of metal, 
since the lone pair electrons may promote a certain structural distortion~\cite{N.V.Sidgwick,Livny1998}.
However, the research seeking for novel properties from metallic materials with 
Pb$^{2+}$ or Bi$^{3+}$ cations is rather limited~\cite{KimberPRL2009,ChengPRB2009,Goodenough2015}.

In this paper, we report the synthesis of a Pb-based perovskite PbMoO$_3$ 
crystalizing in the cubic structure, despite the presence of Pb$^{2+}$ cations.
While this material shows an orthorhombic distortion with 
coherent octahedral tiltings at low temperature,
it retains a metallic behavior down to 0.1~K.
We found an unusual temperature dependence of the electrical resistivity $\rho$
and a large specific heat $C_P$ accompanied by a peak in $C_P/T^3$.
These unusual behaviors are likely originated  from a low-energy vibrational mode 
induced by
the incoherent off-centering of lone pair Pb$^{2+}$ cations as suggested from structural 
refinements. 
The weak $T$ dependence of $\rho$ is presumably understood in terms of 
incoherent transport driven by the formation of the polaron-like conduction.

\section{Experimental}
Polycrystalline samples of Pb$_{1-x}$Sr$_x$MoO$_3$ ($x=0.0, 0.25, 0.5, 0.75$),
were synthesized by a high-pressure and high-temperature technique.
Stoichiometric mixtures of PbO, SrO, and MoO$_2$ were reacted at 1000~$^{\circ}$C and 7~GPa for 30~min
using a multianvil press. 
SrMoO$_3$ ($x=1$) was synthesized by heating SrMoO$_4$ 
(prepared by a solid-state reaction) at 1000~$^{\circ}$C for 12~h 
in H$_2$/Ar flow as reported in Ref.~\onlinecite{HayashiMRB1979}. 

X-ray powder-diffraction (XRD) experiments were performed 
with Cu $K\alpha$ radiation at room temperature (RT).
For an additional structural study of PbMoO$_3$, synchrotron X-ray powder diffraction (SXRD) experiments
were performed at RT on the BL02B2 beam line at SPring-8.
The wavelength of the incident beam was $\lambda= 0.42089$~\AA. 
Neutron diffraction experiments of PbMoO$_3$ were performed at RT and 5 K 
using the high resolution powder diffractometer BT-1 at the NIST center for Neutron Research. 
Incident neutrons of wavelength $\lambda=1.5398$~\AA\, monochromated by 
vertical-focused Cu (311) monochromator was used.
The structural refinements were performed using the FullProf and RIETAN-FP softwares~\cite{Carvajal1993,Izumi2007}.
The $x=0$ sample was also characterized using transmission electron microscopy (TEM). 
The data were acquired on an aberration-corrected FEI Titan 80-300 microscope at 300 kV.
The chemical composition was analyzed by an energy dispersive X-ray spectroscopy (EDS).
The valence was checked by the X-ray absorption near edge structure (XANES) spectroscopy at
the Mo $K$-edge. The spectra of PbMoO$_3$, SrMoO$_3$, MoO$_2$ and MoO$_3$ were measured at RT on 
the BL01B1 beamline at SPring-8. 
The XANES spectra were recorded in a transmittance mode, 
using Si (111) and Si (311) double crystal monochromator.

The specific heat ($C_P$) and dc magnetic susceptibility ($M/H$) were measured, 
respectively, with
a commercial calorimeter (Quantum Design, PPMS) and a SQUID magnetometer 
(Quantum Design, MPMS). The electrical resistivity $\rho$ was measured by means of
a standard four-probe method using rectangular samples cut out from pellets.
Gold wires were attached to samples with silver paste,
and the samples were then cooled down to 0.1~K using an adiabatic demagnetization refrigerator installed in PPMS.
Note that the density of the polycrystalline sample pellet is 8.7~g/cm$^3$ for PbMoO$_3$, 
7.4~g/cm$^3$ for Pb$_{0.5}$Sr$_{0.5}$MoO$_3$, and 5.4~g/cm$^3$ for SrMoO$_3$.
These values correspond to 89--97\% of the calculated ones estimated from the RT crystal structures,
suggesting that the difference in density provides only small effect on the absolute values of 
$\rho$ in these compounds.

\section{Results and discussion}
\label{result}
\begin{figure}[t]
\begin{center}
 \includegraphics[width=0.45\textwidth]{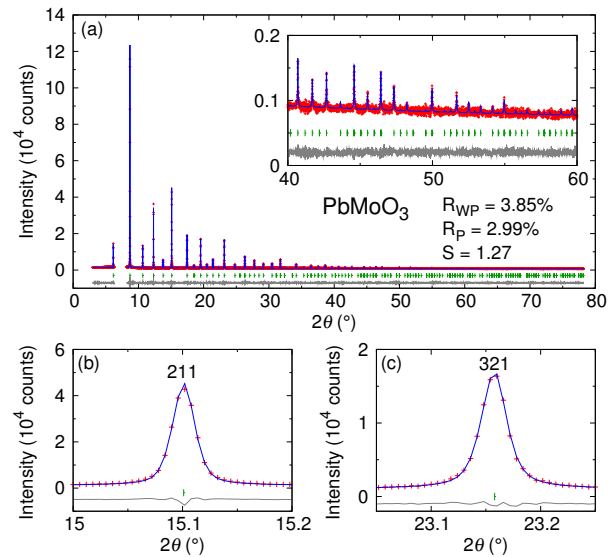}
\caption{
(Color online)
(a)Synchrotron patterns of PbMoO$_3$ measured at RT,
demonstrating the cubic perovskite structure ($Pm\bar{3}m$).
The inset shows the enlarged plot in a high angle region.
(b) and (c) are 211 and 321 reflections, respectively,
representative peaks with $h\ne k$, $\ne l$, and $l\ne 0$,
showing the absence of splitting and shoulder associated with 
any tetragonal or other symmetry lowering distortions.
Observed and refined data are shown by cross and solid curves, 
and vertical bars represent positions of the Bragg reflections.
The difference between the experimental and theoretical data is plotted by 
the dashed curves at the bottom.
Excluded regions in (a) include contributions from small amount of 
impurities such as PbMoO$_4$.
}
\label{fig.1}
\end{center}
\end{figure}

\begin{figure}[t]
\begin{center}
 \includegraphics[width=0.45\textwidth]{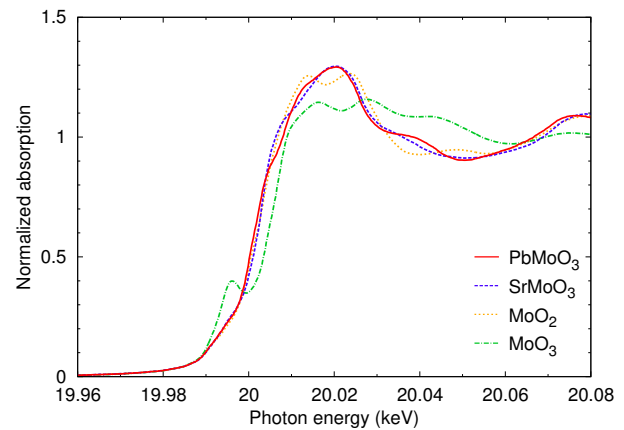}
\caption{
(Color online)
Mo $K$ shell XANES spectra of 
PbMoO$_3$ (solid line), SrMoO$_3$ (dashed line), MoO$_2$ (dotted line), and MoO$_3$ (dashed dotted line).
}
\label{fig.2}
\end{center}
\end{figure}

\begin{table}[t]
\begin{center}
 \caption{Structure parameters for PbMoO$_3$ refined by Rietveld analysis of  
 the synchrotron X-ray and neutron diffraction data. 
 For each model, the Mo atom was positioned at the origin.
 $U_{\mathrm{iso}}$ represents the isotropic atomic displacement parameter. 
 Numbers in parentheses indicate one standard deviation in the value.
 }
 \begin{tabular*}{0.46\textwidth}{@{\extracolsep{\fill}}llll}
  \hline\hline
                  & X-ray (RT)      & neutron (RT) & neutron (5 K)        \rule{0mm}{4mm}   \\ \hline
   \multicolumn{4}{l}{{Cell parameters and positions}}                 \rule{0mm}{4.5mm} \\ 
   \,Space group  & $Pm\bar{3}m$   & $Pm\bar{3}m$& $Imma$                \rule{0mm}{3.5mm} \\
   \,$a$ (\AA)    & 3.999(3)       & 3.9986(1)   & 5.6348(2)           \rule{0mm}{3.5mm} \\
   \,$b$ (\AA)    &                &             & 7.9680(3)           \rule{0mm}{3.5mm} \\
   \,$c$ (\AA)    &                &             & 5.6508(2)           \rule{0mm}{3.5mm} \\
   \,Pb $x$       & 0.5            & 0.5         & 0.0                 \rule{0mm}{4mm}   \\
   \,Pb $y$       & 0.5            & 0.5         & 0.25                \rule{0mm}{3.5mm} \\
   \,Pb $z$       & 0.5            & 0.5         & 0.5008(17)          \rule{0mm}{3.5mm} \\
   \,O1 $x$       & 0.5            & 0.5         & 0.0                 \rule{0mm}{4mm}   \\
   \,O1 $y$       & 0.0            & 0.0         & 0.25                \rule{0mm}{3.5mm} \\
   \,O1 $z$       & 0.0            & 0.0         & 0.0259(12)          \rule{0mm}{3.5mm} \\
   \,O2 $x$       &                &             & 0.25                \rule{0mm}{4mm}   \\
   \,O2 $y$       &                &             & -0.0128(5)          \rule{0mm}{3.5mm} \\
   \,O2 $z$       &                &             & 0.25                \rule{0mm}{3.5mm} \\
   \multicolumn{4}{l}{$U_{\mathrm{iso}}$ ($10^{-2}$\AA$^2$)}\rule{0mm}{4.5mm}   \\
   \,Pb          & 2.08(6)         & 2.58(5)     & 1.2(2)              \rule{0mm}{4mm}   \\
   \,Mo          & 0.30(6)         & 0.64(4)     & 0.36(4)             \rule{0mm}{3.5mm} \\
   \,O1          & 0.8(3)          & 0.99(6)     & 0.6(2)              \rule{0mm}{3.5mm} \\
   \,O2          &                 &             & 0.42(8)             \rule{0mm}{3.5mm} \\
  \hline\hline
 \end{tabular*}
 \label{table_1}
\end{center}
\end{table}
Figure~\ref{fig.1} shows the powder SXRD pattern of PbMoO$_3$ recorded at RT.
Observed peaks are indexed in a cubic unit cell, 
without any splitting and shoulder indicating symmetry lowering
(see, for example, 211 and 321 reflections in Figs.~\ref{fig.1}(b) and \ref{fig.1}(c)).
This result demonstrates the absence of any distortions from the cubic symmetry at RT.
The Rietveld refinement assuming the ideal cubic perovskite (space group: $Pm\bar{3}m$)
converged resulting in the lattice parameter of $a=3.999(1)$~\AA\,
and $R$ factors of
$R_{\rm wp} = 3.85$\%, 
$R_{\rm e} = 3.03$\%,
$R_{\rm p} = 2.99$\%, and
$R_{\rm B} = 3.51$\%.
The goodness-of-fit parameter, $S=R_{\rm wp}/R_{\rm e}$, was $S=1.27$,
indicating excellent quality of the fitting.
The solid solution Pb$_{1-x}$Sr$_x$MoO$_3$ was successfully obtained in the entire $x$ range, 
with the lattice constant decreasing linearly with increasing $x$ (Fig.~1 of the Supplemental Material~[\onlinecite{supplement_PMO}]).
These results ensure a continuous change in the $A$-site composition.
Refining the occupancy factor of each atomic site did not improve the overall fit,
indicating the stoichiometric composition of the title compound at least for
the heavy atoms on the $A$ and $B$ sites.
The Mo $K$-edge XANES spectra of PbMoO$_3$ (Fig.~\ref{fig.2}), 
measured together with other Mo compounds, 
indicate that the formal valence in PbMoO$_3$ is Mo$^{4+}$ (thus yielding Pb$^{2+}$),
which is consistent with the result of bond valence sum calculations based on 
the RT structure, 
giving Mo$^{4.08+}$ and Pb$^{1.74+}$. 
EDS experiments verified the compositional ratio between Pb and Mo, for example, 
Pb/Mo = 1.01(6) for the $x=0$ sample.

\begin{figure}[t]
\begin{center}
 \includegraphics[width=0.35\textwidth]{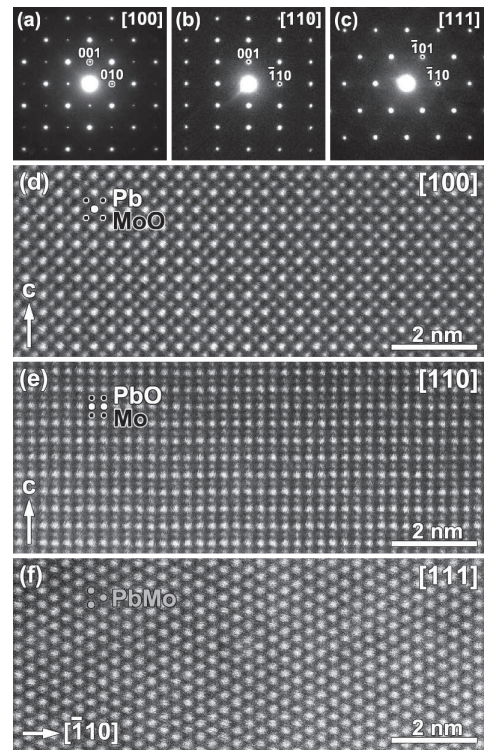}
\caption{
(a)--(c) Electron diffraction patterns and (d)--(f) high resolution 
HAADF-STEM images of PbMoO$_3$ taken at RT along the [100], [110], and [111] axes.
The intensity $I$ in the HAADF-STEM images
is proportional to the average atomic number, $Z$, of the projected atomic column and 
scales as $I\sim Z^{n} (n=1.6-1.9)$. 
}
\label{fig.3}
\end{center}
\end{figure}
Interestingly, a large atomic displacement parameter (ADP) of Pb 
was obtained (isotropic root-mean-square displacement of 0.15(1)\AA), 
suggesting unusually large thermal vibration of Pb ions and/or 
random displacements of Pb away from the ideal position.
Given the presence of stereochemically active 6$s$ electrons of Pb$^{2+}$, 
the obtained cubic structure in PbMoO$_3$ is unusual 
and contrasts markedly with other reported Pb-based perovskites with 
a distortion from the cubic symmetry~\cite{Goodenough2015}.
In order to gain more insight into the RT crystal structure for PbMoO$_3$, 
we performed TEM experiments.
As shown in Figs.~\ref{fig.3}(a)--(c),
electron diffraction patterns can be consistently indexed on the cubic perovskite 
structure with the $Pm\bar{3}m$ symmetry. They do not contain any extra reflections or diffuse 
intensities associated with long or short range ordered structure deformations due to the Pb off-center displacements.
Also, high-angle-annular-dark-field scanning-transmission-electron microscopy (HAADF-STEM) images in 
Figs.~\ref{fig.3}(d)--(f) do not show any notable discrepancy from the ideal cubic perovskite structure. 
No anomaly associated with Pb disorder is observed in any principle directions of 
the cubic structure, [100], [110], and [111].
These results suggest that the Pb off-center displacements, if present, are random at RT,
and the local cooperative structure related to correlated disorder~\cite{KeenNature2015} or 
the formation of polar nano region for relaxor phenomena of Pb-based complex perovskites~\cite{Burns1983,Vakhrushev1989,Hirota2002}
is absent.

Rietveld refinement of the RT neutron diffraction data gave a consistent result compared to the SXRD data,
confirming the full occupancy of the anionic site, 
while at $T=5$~K we found a orthorhombic distortion with additional peaks corresponding to
a $\sqrt{2}\times\sqrt{2}\times2$ cell (Fig.~2 of the Supplemental Material~[\onlinecite{supplement_PMO}]): 
a preliminary SXRD experiment shows the structural phase 
transition appears at $T_{\rm s} \simeq 200$~K.
This superstructure is presumably related to a structural transition as 
observed in SrMoO$_3$ below 125~K involving the softening of a $R_{\rm 25}$ phonon mode 
that results in MoO$_6$ octahedral tilting~\cite{MacquartJSSC2010}.
Indeed, a successful structural refinement at 5~K was conducted assuming 
a low-temperature structural phase analogous to SrMoO$_3$ (space group: $Imma$):
details are given in Tables~\ref{table_1} and the Supplemental Material~[\onlinecite{supplement_PMO}]. 
It is notable that, unlike SrMoO$_3$, the displacement parameter of the Pb site 
is still quite large even at 5 K (root-mean-square displacement of 0.15~\AA\, along the $a$ axis 
and $\sim0.08$~\AA\, along the $b$ and $c$ axes), 
suggesting a strong ``statically'' remaining incoherent Pb off-centering.
It is worth underlining that the difference in $U_{\rm iso}$ with $T$ for Pb between RT and 
5~K is at least three times higher than that for Mo in PbMoO$_3$ or for Sr in SrMoO$_3$~\cite{MacquartJSSC2010},
additionally revealing an unusual thermal behavior possibly related to the dynamic component of lead disorder.
We will discuss the effect of the Pb off-centering on physical properties in the next section. 
As will be shown below, the absence of any anomalies related to the structural transition in $C_P$, $M/H$ and $\rho$
may be related to the associated subtle octahedral tilt of about $4^{\circ}$ along the $a$ axis.

\begin{figure}[t]
\begin{center}
 \includegraphics[width=0.45\textwidth]{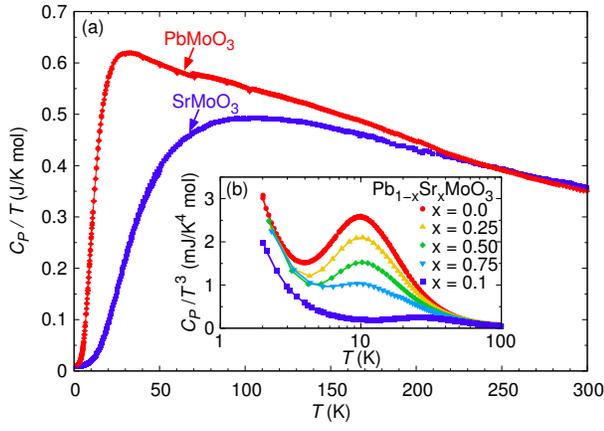}
\caption{
(Color online)
Temperature dependence of $C_P/T$ for powder samples of PbMoO$_3$ and SrMoO$_3$.
Large difference of $C_P/T$ emerges below 150~K, suggesting 
the contribution of a low-energy vibrational mode in PbMoO$_3$.
The inset represents the $C_P/T^3$ vs $T$ plot for Pb$_{1-x}$Sr$_x$MoO$_3$.
}
\label{fig.4}
\end{center}
\end{figure}

It is expected that 
the off-centered Pb$^{2+}$ derived from lone pair electrons in PbMoO$_3$ 
may provide essential influence on physical properties. 
In fact, $C_P/T$ of PbMoO$_3$ is much larger than that of SrMoO$_3$ below 150~K.
Moreover, the $C_P/T^3$ vs. $T$ plot for PbMoO$_3$ clearly exhibits 
a peak centered at about 10~K, which is however 
absent in SrMoO$_3$ (see the inset of Fig.~\ref{fig.4}).
This low-temperature peak cannot be explained in terms of 
the structural phase transition of PbMoO$_3$ and
the simple Debye law with $C_P(T) \propto T^3$,
and hence implies a significant contribution of a low-energy vibrational mode
to the specific heat~\cite{RamirezPRL1998,T.H.K.Barron}. 
It is also seen that Sr-for-Pb substitution does not alter 
the $C_P/T^3$ peak position (temperature), while reducing only its intensity.
It can thus be deduced that this peak originates from the incoherent displacement of Pb ions by
the $6s$ lone-pair electrons, revealed by the large ADP of Pb,
since conventional disorder is known to shift the peak temperature~\cite{R.O.Pohl1985,SafarikPRL2006,MelotPRB2009}.
The $C_P/T^3$ peak for PbMoO$_3$ 
can be roughly reproduced by the Einstein specific heat
with an Einstein temperature $\varTheta_{\rm E}\simeq50$~K
which corresponds to the value estimated from the ADP value of Pb at RT~\cite{supplement_PMO}.
An upturn of $C_P/T^3$ observed in Pb$_{1-x}$Sr$_x$MoO$_3$ systems below 5 K is ascribed to
the electronic specific heat $\gamma T$, and
we obtained $\gamma = 9.2(1)$~mJ/K$^2$-mol for PbMoO$_3$ and $\gamma = 7.6(1)$~mJ/K$^2$-mol for SrMoO$_3$, 
the latter being in agreement with the previous estimation~\cite{Ikeda2000,NagaiAPL2005}. 
The Wilson ratio $R_{\rm W}\equiv\pi^2k_{\rm B}^2\chi_0/(3\mu_{\rm B}^2\gamma)$ is 
obtained as 1.8 and 2.2 for PbMoO$_3$ and SrMoO$_3$, respectively,
using the  Pauli paramagnetic susceptibility $\chi_0$ after diamagnetic correction 
for PbMoO$_3$ and $\chi_0$ from the report on SrMoO$_3$~\cite{Ikeda2000,supplement_PMO}.
The values of $R_{\rm W}$ for both compounds are close to 2 as expected in 
Fermi liquids within the strong correlation limit~\cite{H.Kronmuller}. 
%

\begin{figure}[t]
\begin{center}
 \includegraphics[width=0.45\textwidth]{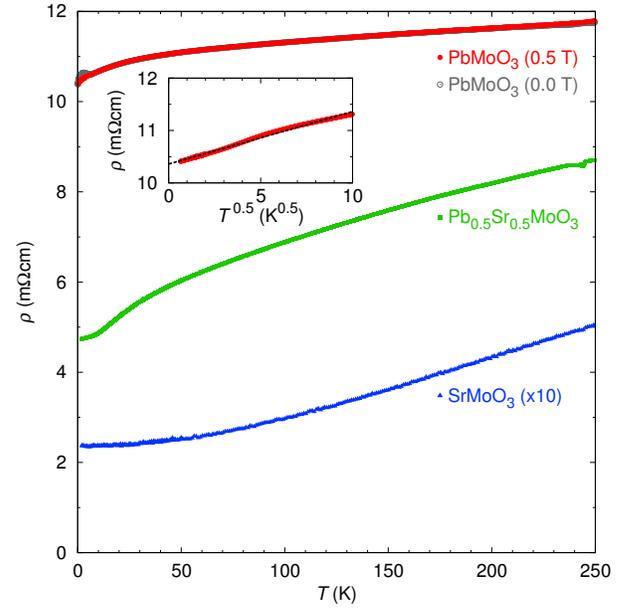}
\caption{
(Color online)
Temperature dependence of $\rho$ for PbMoO$_3$, Pb$_{0.5}$Sr$_{0.5}$MoO$_3$, and SrMoO$_3$.
Here the data for SrMoO$_3$ are multiplied by 10 for clarity.
A tiny difference of the data at 0 and 0.5~T below 7~K for PbMoO$_3$ is attributed to 
small amount of a Pb impurity which is less than $0.1$\% examined from diffraction and 
magnetic susceptibility experiments. 
Other signatures of SC transition or spin density wave have not been observed down to 0.1~K
in PbMoO$_3$.
The inset shows the low-temperature behavior of $\rho$ plotted as a function of $T^{0.5}$ for PbMoO$_3$.
}
\label{fig.5}
\end{center}
\end{figure}
%

Temperature dependence of the electrical resistivity $\rho(T)$ of PbMoO$_3$ (Fig.~\ref{fig.5}) shows 
a metallic behavior down to 0.1~K (apart from a tiny contribution of the superconductivity (SC) 
from the Pb impurity below 7~K).
This SC contribution can be removed by applying a magnetic field of 0.5 T, 
which is above the critical field of the SC transition of Pb.
The resulting $\rho(T)$ is almost identical with the zero-field data,
implying that the magnetoresistance is negligible.
Interestingly, $\rho(T)$ of PbMoO$_3$ exhibits an unusual $T$-sub linear dependence 
in a wide temperature range below 100~K, where the fitting with 
$\rho(T) = A + B T^{\alpha}$ to the data yielded $\alpha \simeq 0.5$ (the inset of Fig.~\ref{fig.5}).
Moreover, the resistive change in temperature is rather small, with a residual resistivity ratio
$\rm{RRR}=\rho(300~\rm{K})/\rho(0.1~\rm{K})=1.1$.
These features are distinct from $\rho(T)$ of typical non-magnetic metals, where 
$\rho(T)\propto T$ or $\propto T^5$ due to the electron-phonon scattering with the weak electron-phonon coupling~\cite{Ziman1,Allen1993}
and/or $\rho(T)\propto T^2$ due to the electron-electron scattering are observed at low temperature~\cite{A.P.Mackenzie2003,Kadowaki1986}.
Note that SrMoO$_3$ shows $\rho(T)\propto T^2$ below 140~K (the main panel of Fig.~\ref{fig.5})
owing to 
the enhanced electron-electron correlation~\cite{NagaiAPL2005}.
We also observed that $\rho(T)\propto T^2$ 
is recovered by substituting Pb ions by Sr ions at low temperatures
for the $x=0.5$ sample.
Therefore, it appears that incoherent off-centered Pb$^{2+}$ ions 
disturb the electrical conduction.
It is known that a weak coupling between optical phonons and conduction electrons 
gives rise to a $T$-super linear dependence with $\rho(T)\propto T^n$ ($n\ge1$)~\cite{Allen1993,Masui2002}.
It is thus possible that the observed polaron-like incoherent transport is mediated by 
strong coupling between conduction electrons and optical phonons of the local low-energy vibrational mode for PbMoO$_3$, 
as proposed theoretically by Millis {\it{et al}.}~\cite{MillisPRB1996-1}
It was suggested that $\rho(T)\propto T^{0.5}$ can appear in 
the critical regime around the crossover from Fermi liquid to polaron behaviors.

\section{Conclusion}
To summarize, we have synthesized a novel lead perovskite PbMoO$_3$ 
using a high-pressure and high-temperature reaction. 
This lead-based compound represents a rare case with 
the $Pm\bar{3}m$ cubic structure at room temperature, 
despite the presence of Pb$^{2+}$ cations with lone pair electrons.
We observed an unusual $T$-sub linear dependence in $\rho(T)$
as well as large specific heat at low temperatures, 
which could be explained in terms of a low-energy vibrational mode
mediated by the incoherent off-centering of Pb$^{2+}$ cations
as experimentally indicated by anomalously large Pb atomic displacements.
Furthermore,
the weak $T$ dependence of $\rho(T)$ implies a polaron-like conduction, mediated by
the strong trapping of conduction electrons by local phonon vibration.

%
\textit{Note added}. 
Recently, we became aware of the work on Pb$_2$Cr$_{1+x}$Mo$_{1-x}$O$_6$
reporting a synthesis for $x=-1, -2/3, -1/3, 0, 1/3$ ($x=-1$ means PbMoO$_3$),
yet with no structural refinement (thus no indication of Pb off-centering) 
or detailed physical properties~\cite{ZhaoJSSC2017}.

\section*{Acknowledgment}
This work was supported by CREST and
JSPS KAKENHI Grants No. 25249090, No. 24248016, No. 26106514, and No. 26400336.
The synchrotron radiation experiments, performed at the BL02B2 of SPring-8, were supported by 
the approval of the Japan Synchrotron Radiation Research Institute (JASRI) (Proposal No. 2014B1360).
%

\pagebreak
\widetext
\begin{center}
\textbf{\large Supplemental Material: Cubic lead perovskite PbMoO$_3$ with anomalous metallic behavior}
\end{center}
\setcounter{equation}{0}
\setcounter{figure}{0}
\setcounter{table}{0}
\setcounter{page}{1}
\setcounter{section}{0}
\makeatletter
\renewcommand{\theequation}{S\arabic{equation}}
\renewcommand{\thefigure}{S\arabic{figure}}
\renewcommand{\bibnumfmt}[1]{[S#1]}
\renewcommand{\citenumfont}[1]{S#1}

\section{Abstract}
In this supplemental material, 
we describe the synthesis of solid solutions of Pb$_{1-x}$Sr$_x$MoO$_3$ ($x = 0.0, 0.25, 0.5, 0.75, 1.0$) 
with these powder X-ray diffraction (XRD) data.
We also show the result of neutron diffraction measurements
and the analysis of the specific heat $C_P$ as well as the result of magnetic susceptibility $M/H$ 
for the $x=0$ sample of PbMoO$_3$.

\section{Synthesis of Pb$_{1-x}$Sr$_x$MoO$_3$}
\begin{figure}[h]
\begin{center}
 \includegraphics[width=0.45\textwidth]{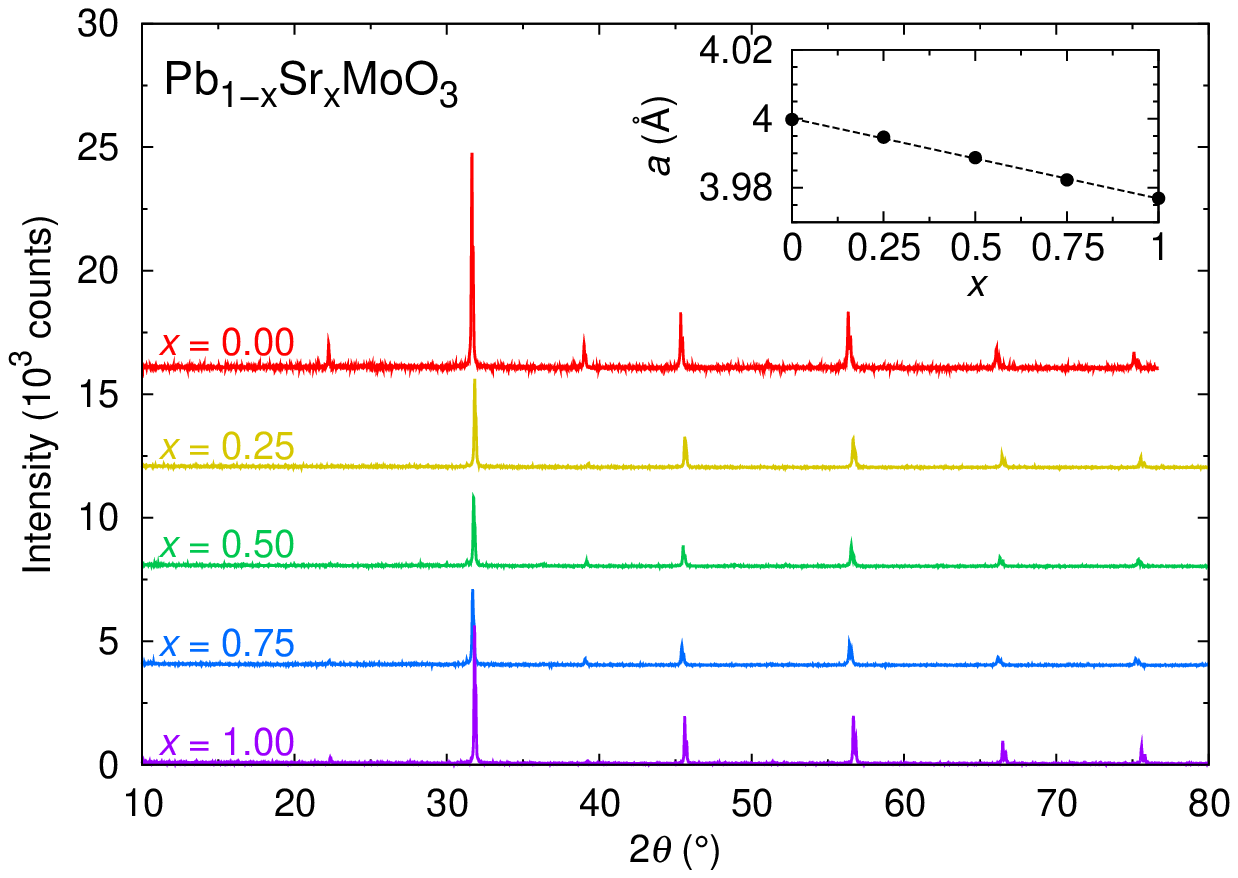}
\caption{
(Color online)
XRD patterns of Pb$_{1-x}$Sr$_x$MoO$_3$ ($x=0.0, 0.25, 0.5, 0.75, 1.0$) at RT.
Inset is the lattice parameter $a$ plotted as a function of $x$.
}
\label{fig.S1}
\end{center}
\end{figure}

Samples of Pb$_{1-x}$Sr$_x$MoO$_3$ were obtained by 
the same method for the synthesis of PbMoO$_3$. 
XRD patterns of these samples
obtained through a laboratory X-ray machine at room temperature (RT)
represent the same pattern of PbMoO$_3$ (Fig.~\ref{fig.S1}),
indicating that the samples crystalize into the cubic structure in the entire $x$ range.
The estimated lattice parameters linearly decrease with increasing $x$ (the inset of Fig.~\ref{fig.S1}),
which ensures the stoichiometric mixture of the samples.
It is observed that impurity peaks such as PbMoO$_4$ and Pb appear in the XRD patterns,
however those concentrations are estimated to be small enough ($<1\%$).
\begin{figure}[h]
\begin{center}
 \includegraphics[width=0.45\textwidth]{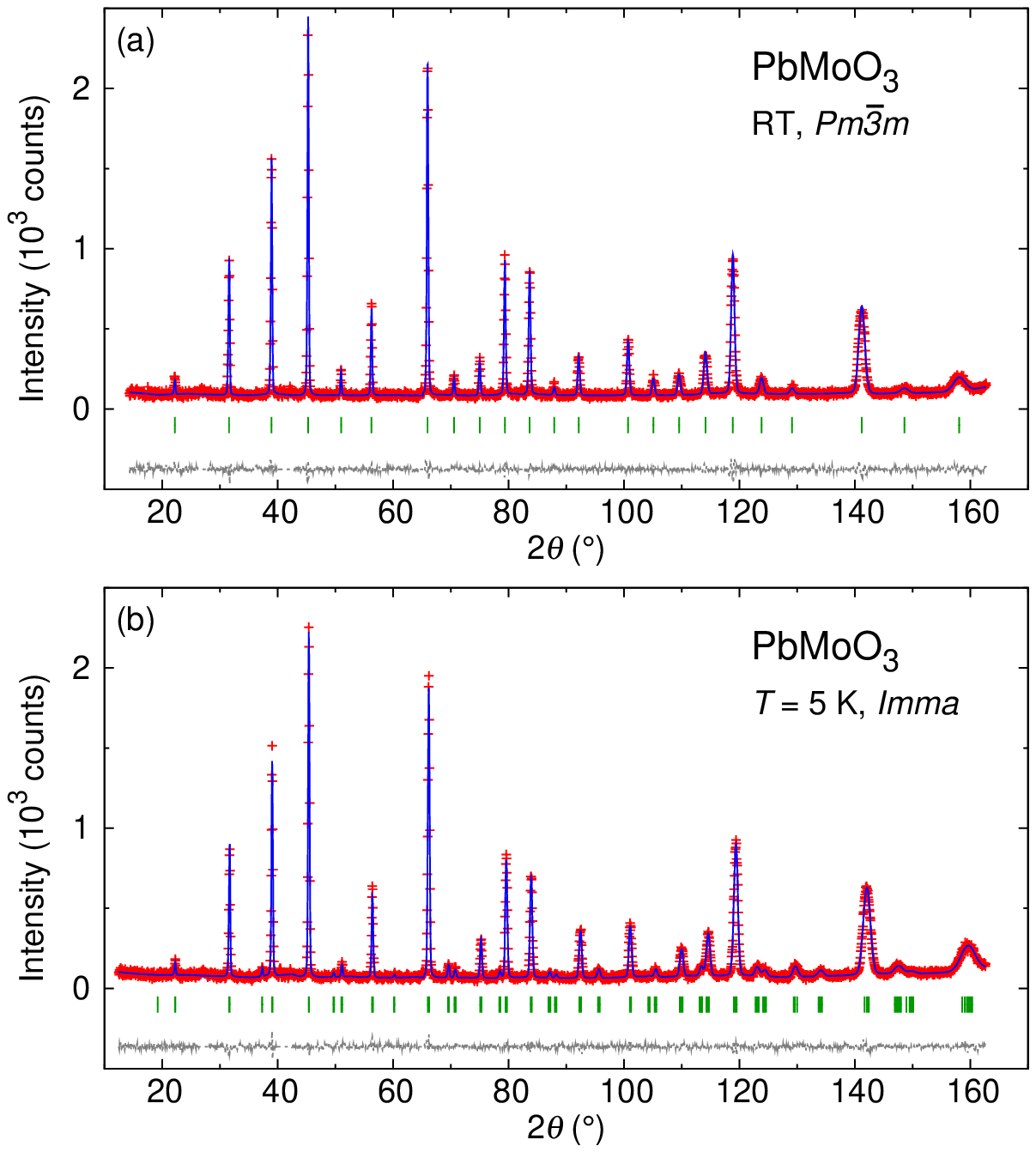}
\caption{
(Color online)
Neutron diffraction patterns of PbMoO$_3$ measured at (a) RT and (b) 5~K.
Observed and refined data are shown by cross and solid curves, respectively, 
and vertical bars represent positions of the Bragg reflections.
The difference between the experimental and theoretical data is plotted by 
the dashed curves at the bottom.
}
\label{fig.S2}
\end{center}
\end{figure}

Neutron diffraction measurements on the $x=0$ sample at 5~K
show an additional peak or splitting of peaks from the $Pm\bar{3}m$ cubic symmetry observed at RT 
(e.g., around 70$^\circ$ in Fig.~\ref{fig.S2}).
These peaks are not magnetic in the origin, because magnetic susceptibility experiments 
did not show any signature of magnetic phase transition.
We therefore analyzed the low-temperature crystal structure,
assuming the $Imma$ structure which is 
the orthorhombic space group (SG) of the highest symmetry: this choice is reasonable because 
other orthorhombic SGs of lower symmetry, fulfilling the observed $I$-centering, such as 
$Immm$, $Ibam$, $Ibca$, $Imm2$, $Iba2$, $Ima2$, $I222$, and $I2_12_12_1$ are 
incompatible with the extinction conditions.
We also reexamined the RT $Pm\bar{3}m$ structural model of 
the synchrotron XRD analysis by the Rietveld refinement against neutron data.
The resulting curves are shown in Fig.~\ref{fig.S2}, and
the obtained structural parameters are presented in Table~I of the main text.
The $R$ factors of the Rietveld refinement are 
$R_{\rm wp} = 16.3$\%, 
$R_{\rm e}  = 17.8$\%,
$R_{\rm p}  = 19.2$\%, 
$R_{\rm B}  = 2.28$\% for RT, and
$R_{\rm wp} = 15.9$\%, 
$R_{\rm e}  = 16.9$\%,
$R_{\rm p}  = 16.2$\%, 
$R_{\rm B}  = 2.56$\% for 5~K.
The goodness-of-fit parameter, $S=R_{\rm wp}/R_{\rm e}$, was $S=0.915$ and 0.940 for 
RT and 5~K, indicating excellent quality of fittings.

\begin{figure}[t]
\begin{center}
\includegraphics[width=0.60\textwidth]{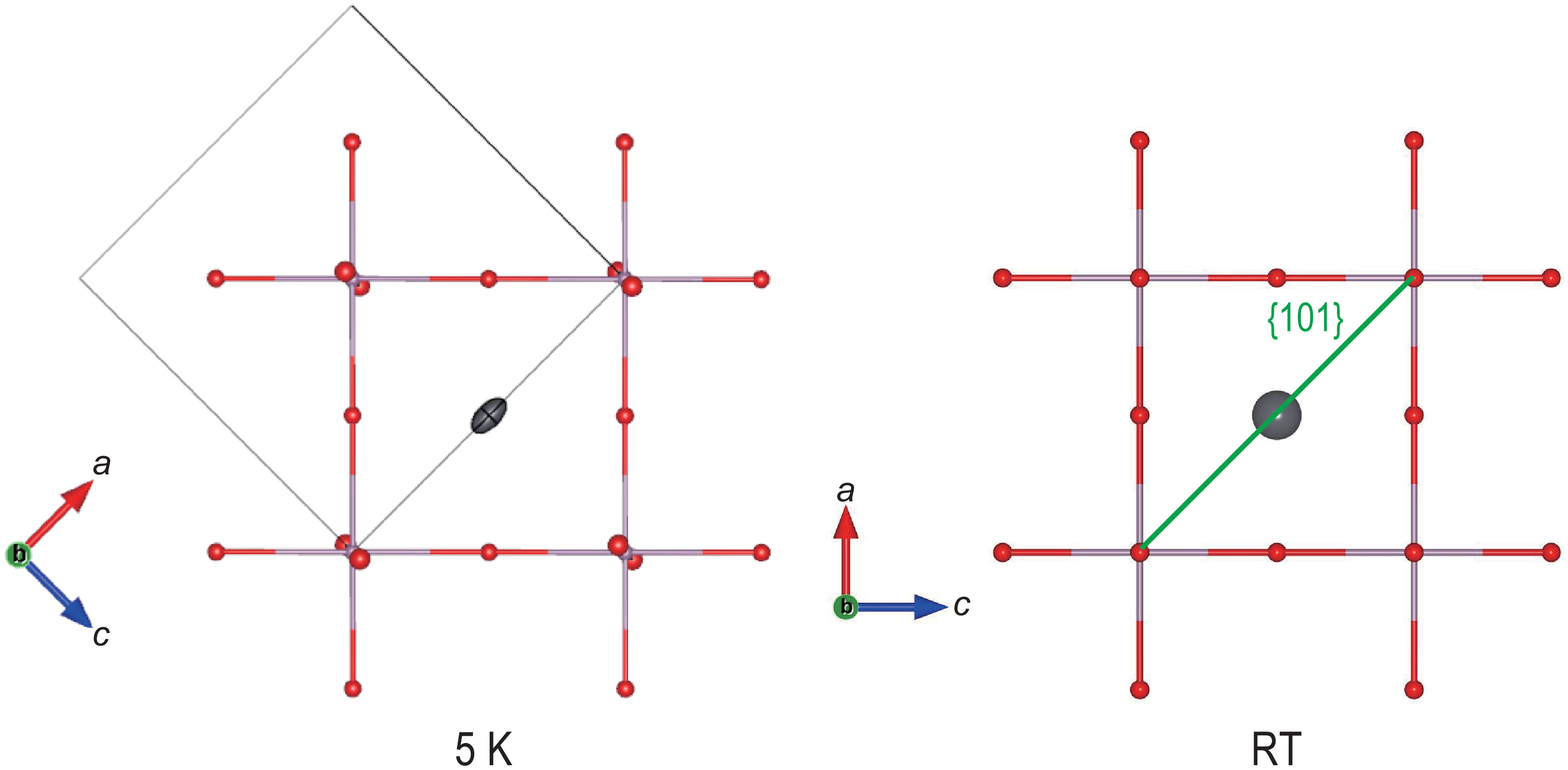}
\caption{
(Color online)
Crystal structures of PbMoO$_3$ at 5~K (left) and RT (right), featuring the Pb environment.
Gray ellipsoid or sphere represent the Pb ion drawn with the anisotropic displacement parameters.
Red and purple spheres represent O and Mo ions, respectively.
A squre lattice, shown by gray lines in the left, represents the unit cell of the $Imma$ structure.
}
\label{fig.S3}
\end{center}
\end{figure}

%
The key feature of the low temperature $Imma$ structe in PbMoO$_3$ lies in 
the elongated shape of the Pb anisotropic displacement parameters along the $a$ axis
that corresponds to the [101] direction of the high temperature cubic unit cell (Fig.~\ref{fig.S3}).
This result is also supported by using difference Fourier syntheses and maximum entropy method (MEM) calculations.
These analyses clarify that the Pb nuclear density is homogeneously elongated along the $a$ axis 
without any maximum density on each side of the $4e$ Wyckoff site.
\section{Analysis of specific heat}
\begin{figure}[t]
\begin{center}
 \includegraphics[width=0.45\textwidth]{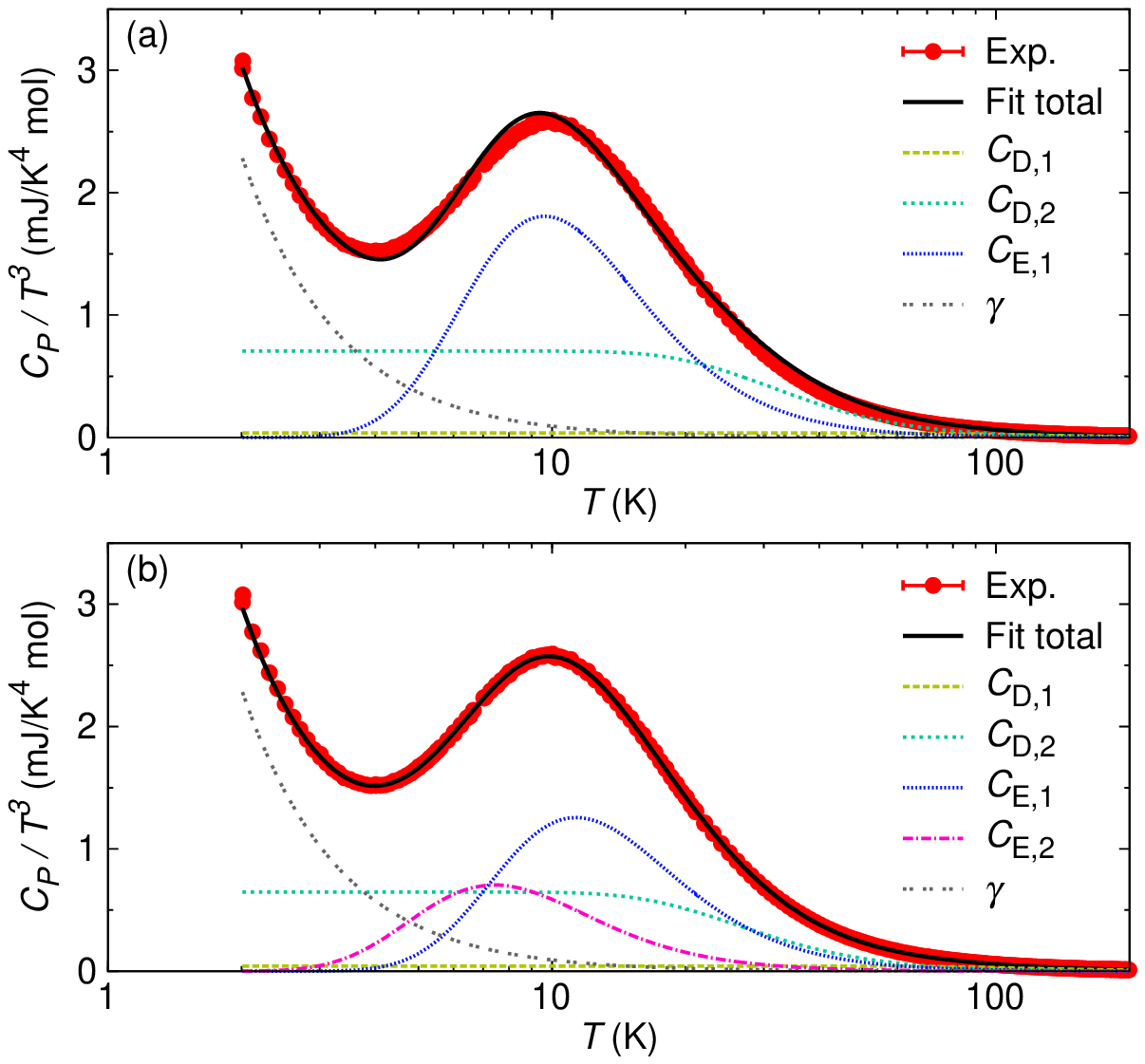}
\caption{
(Color online)
Results of fitting of the specific heat data shown in the inset of Fig.~4 
of the main text. 
In (a), two Debye and one Einstein terms were used for the fitting with Eq.~(\ref{eq.2}),
while in (b), two Debye and two Einstein terms were used.
}
\label{fig.S4}
\end{center}
\end{figure}
\begin{table}[t]
\begin{center}
 \caption{Parameters obtained by fitting of the specific heat data for PbMoO$_3$.
 Here, the Debye and Einstein modes of Eq.~(\ref{eq.1}) are represented as D$i$ ($i=1,2$) and E$j$ ($j=1,2$), respectively.
 }
 \begin{tabular*}{0.46\textwidth}{@{\extracolsep{\fill}}ccc}
  \hline\hline
   \multicolumn{3}{l}{Two Debye and one Einstein model [Fig.~\ref{fig.S4}(a)]}\rule{0mm}{4.5mm}   \\
   \,Mode         & $\varTheta_{{\rm{D}},i}$, $\varTheta_{{\rm{E}},j}$ (K)     & $f_{{\rm{D}},i}$, $f_{{\rm{E}},j}$   \rule{0mm}{4mm}   \\ \hline
   \,D1           & 560                          & 9.92                \rule{0mm}{3.5mm} \\
   \,D2           & 154                          & 3.99                \rule{0mm}{3.5mm} \\
   \,E1           & 47.4                         & 1.09                \rule{0mm}{3.5mm} \\
   \multicolumn{3}{l}{Two Debye and two Einstein model [Fig.~\ref{fig.S4}(b)]}\rule{0mm}{6.5mm}   \\
   \,Mode         & $\varTheta_{{\rm{D}},i}$, $\varTheta_{{\rm{E}},j}$ (K)     & $f_{{\rm{D}},i}$, $f_{{\rm{E}},j}$    \rule{0mm}{4mm}   \\ \hline
   \,D1           & 560                          & 11.4                \rule{0mm}{3.5mm} \\
   \,D2           & 129                          & 2.15                \rule{0mm}{3.5mm} \\
   \,E1           & 55.9                         & 1.23                \rule{0mm}{3.5mm} \\
   \,E2           & 36.9                         & 0.19                \rule{0mm}{3.5mm} \\
  \hline\hline
 \end{tabular*}
 \label{table_S1}
\end{center}
\end{table}
In order to analyze the anomalous behavior of the $C_P(T)/T^3$ peak observed in PbMoO$_3$,
we have fitted the data as follows.

The phonon contribution to the low temperature specific heat, $C_{\rm{ph}}(T)$, can be modeled as 
a sum of Debye and Einstein terms~\cite{RamirezPRL1998_SM,T.H.K.Barron_SM}, 
$C_{{\rm{D}},i}(T)$ and $C_{{\rm{E}},j}(T)$ $[i = 1,2,...; j = 1,2,...]$,
i.e.,
\begin{alignat}{1}
  C_{\rm{ph}}(T) = \sum_{i} f_{{\rm{D}},i}C_{{\rm{D}},i}(T) + \, \sum_{j} f_{{\rm{E}},j}C_{{\rm{E}},j}(T),
  \label{eq.1}
\end{alignat}
where $f_{{\rm{D}},i}$ and $f_{{\rm{E}},j}$ are fitting parameters 
representing the number of the Debye and Einstein modes for the $i$- and $j$-th contributions.
The sum of these coefficients represents the total number of lattice dynamical modes in the unit cell, i.e., 
$\sum_{i} f_{{\rm{D}},i} + \sum_{j} f_{{\rm{E}},j} = 15$ for PbMoO$_3$.
$C_{{\rm{D}},i}(T)$ and $C_{{\rm{E}},j}(T)$ represent 
the contribution of
Debye phonons and Einstein phonons to the specific heat, expressed as
\begin{alignat}{2}
&C_{{\rm{D}},i}(T) = 3\,R \Bigl(\frac{T}{\varTheta_{{\rm{D}},i}} \Bigr)^3 
                \int_0^{\varTheta_{{\rm{D}},i}/T} \frac{e^{x} x^{4} \,dx}{(e^{x}-1)^2}, 
                \label{eq.CD}\\
&C_{{\rm{E}},j}(T) =   R 
                \frac{e^{\varTheta_{{\rm{E}},j}/T}}{(e^{\varTheta_{{\rm{E}},j}/T}-1)^2} 
                \Bigl(\frac{\varTheta_{{\rm{E}},j}}{T} \Bigr)^2,
                \label{eq.CE}
\end{alignat}
where $R$ is the gas constant, and $\varTheta_{{\rm{D}},i}$ and $\varTheta_{{\rm{E}},j}$ 
are fitting parameters representing Debye and Einstein temperatures, respectively.
We tentatively used these formulations for the specific heat analysis for PbMoO$_3$,
in reference to other systems~\cite{RamirezPRL1998,KeppensNature1998,Yamamura2002,TachibanaPRB2009,MelotPRB2009_SM,Takatsu2007,T.H.K.Barron}.

The least-squares fitting was carried out using the data between 2 and 100 K,
and the relation 
\begin{alignat}{1}
  C_{P}(T) = \gamma T + C_{\rm{ph}}(T),
  \label{eq.2}
\end{alignat}
with taking into account the electronic specific heat constant $\gamma=9.2(1)$~mJ/K$^{4}$ mol 
for PbMoO$_3$.
In the primary fit, we considered two Debye and one Einstein modes ($i=2$, $j=1$)
and obtained reasonable fitting results reproducing experimental features 
as shown in Fig~\ref{fig.S4}(a).
It is seen that the $C_P/T^3$ peak is reproduced by the Einstein specific heat with $\varTheta_{{\rm{E}},1}\simeq50$~K.
This value agrees well with the estimation of the Einstein temperature $\varTheta_{{\rm{E}}}=58$~K 
from the isotropic atomic displacement parameter $U_{\rm iso}$ of Pb at RT
for the relation between $U_{\rm iso}$ and $\varTheta_{{\rm{E}}}$~\cite{T.M.Tritt}, i.e., 
$U_{\rm iso} = h^2T/4\pi m k_{\rm B}\varTheta_{{\rm{E}}}^2$,
where $h$ and $k_{\rm B}$ are the Planck and Boltzmann constants, respectively, and $m$ is the atomic mass.
This implies the close relation between the local vibrational mode mediated by incoherent off-centered Pb$^{2+}$ cations and 
large low-temperature specific heat featured by the $C_{P}/T^3$ peak at about 10~K.
The obtained fitting parameters are listed in Table~\ref{table_S1}.
Note that the high-temperature part from one Debye contribution with $\varTheta_{{\rm{D}},1}=560$~K 
provides a reasonable value for high-temperature mode in oxides which mainly contributes to the data for $T>100$~K.
It is worth noting that the model with two Debye and two Einstein modes 
can slightly improve the fitting as shown in Fig.~\ref{fig.S4}(b),
and the obtained fitting parameters are listed in Table~\ref{table_S1}.
This result suggests that a low-energy optical mode of PbMoO$_3$ is weakly
dispersive.

\section{Magnetic susceptibility of PbMoO$_3$}
\begin{figure}[h]
\begin{center}
 \includegraphics[width=0.45\textwidth]{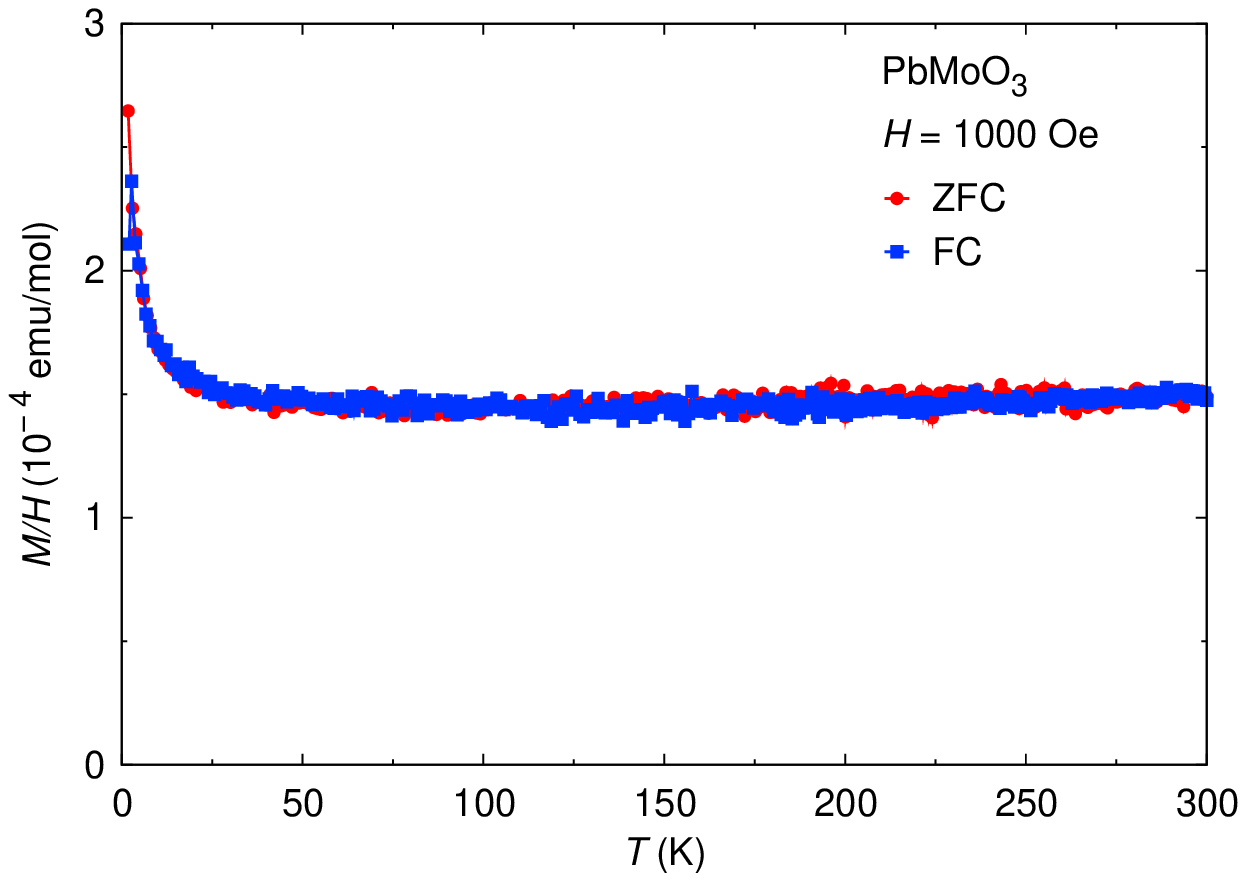}
\caption{
(Color online)
Temperature dependence of the magnetic susceptibility $M/H$ of the polycrystalline 
sample of PbMoO$_3$ at $H=1000$~Oe.
Note, 1~emu = 10$^{-3}$~A m$^{2}$.
}
\label{fig.S5}
\end{center}
\end{figure}

The temperature dependence of $M/H$ of PbMoO$_3$ measured at $H=1000$~Oe 
does not show any signature of the phase transition and hysteresis between zero-field 
cooling and field cooling runs (Fig.~\ref{fig.S5}). 
A tiny diamagnetic signal accompanied by the superconducting transition of Pb impurities
was observed in low field below 7~K, however its amount is less than 1\% in our samples.
Note that in particular, the estimated impurity amount for the sample used in the resistivity measurements 
is less than 0.1\%.
In accordance with a slight enhancement of Pauli paramagnetic susceptibility,
$M/H$ of PbMoO$_3$ is $T$-independent in a wide temperature range above 20~K, while 
the Curie-type increase below 20~K corresponds to small amount (about 0.1\%) of free impurity spins.
The Pauli paramagnetic susceptibility is estimated as $\chi_0=2.2\times10^{-4}$~emu/mol
after the correction of the diamagnetic susceptibility of $\chi_{\rm dia} = -8.1\times10^{-5}$~emu/mol. 
This value of $\chi_0$ is similar to the value reported in SrMoO$_3$~\cite{Ikeda2000_SM,Zhang2006,Zhao2007}.

\end{document}